\documentclass[aps,prd,12pt,amssymb]{revtex4}

\begin{document}

\baselineskip=0.60cm

\newcommand{\ini}{\begin{equation}}
\newcommand{\fin}{\end{equation}}
\newcommand{\inir}{\begin{eqnarray}}
\newcommand{\finr}{\end{eqnarray}}
\newcommand{\inif}{\begin{figure}}
\newcommand{\finf}{\end{figure}}
\newcommand{\bc}{\begin{center}}
\newcommand{\ec}{\end{center}}

\def\ol{\overline}
\def\pa{\partial}
\def\ra{\rightarrow}
\def\ts{\times}
\def\df{\dotfill}
\def\bs{\backslash}
\def\dg{\dagger}
\def\la{\lambda}

$~$

\hfill DSF-22/2005

\vspace{1 cm}

\title{QUARK-LEPTON SYMMETRY AND COMPLEMENTARITY}

\author{D. Falcone}

\affiliation{Dipartimento di Scienze Fisiche,
Universit\`a di Napoli, Via Cintia, Napoli, Italy}

\begin{abstract}
\vspace{1cm}
\noindent
We argue that the difference between the observed approximate quark-lepton
complementarity and the theoretical prediction based on realistic quark-lepton
symmetry within the seesaw mechanism may be adjusted by means of a triplet
contribution in the seesaw formula.
\end{abstract}

\maketitle

\newpage

\section{Introduction}

It is well known that the (type I) seesaw mechanism \cite{ss} is able to
explain the smallness of neutrino mass.
In fact, for a single fermion generation, the effective neutrino mass
$m_{\nu}$ is given by $m_{\nu} \simeq (m_D/m_R)m_D$, where the Dirac
mass $m_D$ is of the order of the quark (or charged lepton) mass, and
the right-handed Majorana mass $m_R$ is of the order of the unification
or intermediate mass scale. As a result, the effective mass comes out
very small with respect to the Dirac mass.

The (type I) seesaw mechanism may as well explain the existence of some
large lepton mixings \cite{smir}.
In fact, for three fermion generations, the effective neutrino mass matrix
$M_{\nu}$ is given by the formula
\ini
M_{\nu} \simeq M_D^T M_R^{-1} M_D,
\fin
so that large neutrino mixing can be generated from a nearly diagonal $M_D$
by means of a strong mass hierarchy or large offdiagonal elements in $M_R$
\cite{smir,js1,afm,df}.
A small contribution to lepton mixing from the charged lepton mass matrix
$M_e$ is also expected and could be important to understand the deviation
from maximal mixing \cite{js2,gt}.

On the other hand, the triplet contribution to the seesaw formula, leading
to the so called type II seesaw mechanism \cite{ss2}, is probably
present (see for example \cite{bsv}).
The type I seesaw mechanism is based on the introduction, within the
standard model, of three heavy right-handed neutrinos. However, small
neutrino masses can be generated also by the inclusion of a heavy Higgs
triplet \cite{ss2,ma}. In this case, the neutrino mass matrix is given by
$M_{\nu}=M_L=Y_L v_L$, where $Y_L$ is a Yukawa matrix and $v_L$ is the
v.e.v. of the triplet, which can be written as $v_L=\gamma v^2/m_T$,
with $v$ the v.e.v. of a standard Higgs doublet, $m_T$ the triplet mass,
and $\gamma$ a coefficient related to the coupling between the doublet
and the triplet. Then, $v_L$ is small with respect to $v$, but
large mixing in $M_{\nu}$ is achieved by hand. Instead, from the type I
seesaw formula we get $M_{\nu} \simeq Y_D^T M_R^{-1} Y_D v^2$, so
that large mixing can be generated from the structure of both matrices
$Y_D$ and $M_R$.
Therefore, we can write the type II seesaw formula by adding to
the usual type I term the triplet (or type II) term, so that
\ini
M_{\nu} \simeq M_D^T M_R^{-1} M_D + M_L.
\fin
The triplet contribution by alone can explain the smallness of neutrino
mass but not the existence of large lepton mixings. Nevertheless, it
can produce important effects on such mixings within the type II seesaw.

It has also been suggested that the generation of maximal mixings by
means of the type I seesaw mechanism could be natural \cite{hm}.
Then, the realistic quark-lepton symmetry \cite{gj} is not consistent with
the approximate quark-lepton complementarity which is observed in
quark and lepton mixings \cite{fp}, that is
\ini
\theta_{12}+\vartheta_{12} \simeq \frac{\pi}{4},
\fin
where $\theta_{12}$ is the 1-2 quark mixing angle (the Cabibbo angle)
and $\vartheta_{12}$ is the 1-2 lepton mixing angle (the solar neutrino
mixing angle). The central point of our paper
is that the triplet contribution in the type II seesaw formula can
indeed correct such a disagreement.
We perform an $SO(10)$ inspired study, where both the realistic
quark-lepton symmetry and the relation between $M_L$ and $M_R$ are
well motivated.  

\section{Framework}

Quark mixing and lepton mixing are quite different from each other.
The quark mixing matrix $V$ is called the CKM matrix, the lepton
mixing matrix $U$ is called the MNS matrix. Now, for the quark mixing
we have \cite{oh}
$$
V_{12}=0.221-0.227
$$
$$
V_{23}=0.039-0.044
$$
$$
V_{13}=0.0029-0.0045
$$
and for the lepton mixing \cite{oh}
$$
U_{12}=0.48-0.62
$$
$$
U_{23}=0.58-0.84
$$
$$
U_{13}<0.22.
$$
Therefore, the largest quark mixing is smaller or at most similar to
the smallest lepton mixing. As we said, the seesaw mechanism could be
the origin of this different behaviour.

Quark and lepton mixings come out from the diagonalization of quark and
lepton mass matrices. A typical expression for the quark mass matrices
is given by \cite{rrrv,akm}
\ini
M_u \simeq
\left( \begin{array}{ccc}
0 & \la^6 & \la^{6} \\
\la^6 & \la^4 & \la^4 \\
\la^{6} & \la^4 & 1
\end{array} \right)~m_t,
\fin
\ini
M_d \simeq
\left( \begin{array}{ccc}
0 & \la^3 & \la^{3} \\
\la^3 & \la^2 & \la^2 \\
\la^{3} & \la^2 & 1
\end{array} \right)~m_b,
\fin
where $\lambda \simeq 0.2$ and coefficients (not written) are close to one.
According to the quark-lepton symmetry, as within the $SO(10)$ model,
$M_e \sim M_d$, $M_D \sim M_u$,
and including the $-3$ factor of Georgi and Jarlskog \cite{gj}, which
gives better the charged lepton masses, we get the following
lepton mass matrices,
\ini
M_e \simeq
\left( \begin{array}{ccc}
0 & \la^3 & \la^{3} \\
\la^3 & -3 \la^2 & \la^2 \\
\la^{3} & \la^2 & 1
\end{array} \right)~m_b,
\fin
\ini
M_D \simeq
\left( \begin{array}{ccc}
0 & \la^6 & \la^{6} \\
\la^6 & -3 \la^4 & \la^4 \\
\la^{6} & \la^4 & 1
\end{array} \right)~m_t,
\fin
where $M_D$ is the Dirac neutrino mass matrix. The $-3$ factor is due to the
contribution of the {\bf 126} representation. The other entries are due to
the {\bf 10} representation. 

It was noted some years ago \cite{df} that in the case of normal hierarchy of
neutrinos the inverse of the neutrino mass matrix has all entries of the same
order of magnitude. Then, by inverting the type I seesaw formula,
$M_{R} \simeq M_D M_{\nu}^{-1} M_D^T$, we can determine the (Majorana) mass matrix
of the right-handed neutrinos, and we yield
\ini
M_R \simeq
\left( \begin{array}{ccc}
\la^{12} & \la^{10} & \la^6 \\
\la^{10} & \la^8 & \la^4 \\
\la^6 & \la^4 & 1
\end{array} \right)~m_R.
\fin
By applying again the direct seesaw formula we obtain the effective neutrino
mass matrix
\ini
M_{\nu}^I \simeq
\left( \begin{array}{ccc}
\la^4 & \la^2 & \la^2 \\
\la^2 & 1 & 1 \\
\la^2 & 1 & 1
\end{array} \right)~\frac{m_t^2}{m_R}.
\fin
These mass matrices form our framework for the type I seesaw mechanism.
In the next section we are going to study its prediction and include the
type II term.

\section{Analysis}

In order to explore the consequences of the foregoing framework, we should
consider the following form of the type I term
\ini
M_{\nu}^I \simeq
\left( \begin{array}{ccc}
\la^4 & \la^2 & \la^2 \\
\la^2 & 1+ \frac{A}{2} & 1- \frac{A}{2} \\
\la^2 & 1-\frac{A}{2}  & 1+\frac{A}{2} 
\end{array} \right)~\frac{m_t^2}{m_R}.
\fin
In fact, the matrix (9) is of course approximate, and may provide different
values for the mixing angles. Nevertheless, we first assume
maximal 2-3 mixing, and then the bimaximal mixing matrix $U_{\nu}$
(that is the 1-2 mixing also maximal) is obtained for $A=\lambda^4$.
When we include the effect of the charged lepton mass matrix on the mixing,
$U_e^{\dg} U_{\nu}$, we get
\ini
U_{12} \simeq \frac{1}{\sqrt2}+\frac{\lambda}{6}
\fin
\ini
U_{23} \simeq \frac{1}{\sqrt2}-\frac{\lambda^2}{\sqrt2}
\fin
\ini
U_{13} \simeq \frac{\lambda}{3 \sqrt2}
\fin
so that $U_{12}$ is out of the experimental range and in particular is
not consistent with the quark-lepton complementarity relation (3).

However, we have also to include the impact of the triplet term
$M_{\nu}^{II}=M_L$, and we would like to take $M_L$ proportional
to $M_R$, so that
\ini
M_{\nu}^{II}=M_L=\frac{m_L}{m_R} M_R.
\fin
This relation is indeed present in left-right models (see for example \cite{rod1}),
including the $SO(10)$ model (see \cite{bsv}). In fact, both $M_R$ and $M_L$
are generated by the {\bf 126} representation. 
We set $m_{\nu}^{II}/m_{\nu}^{I}=k$, the ratio between the overall
scales or the type I and type II terms,
\ini
k=\frac{m_L m_R}{m_t^2}=\gamma ~\frac{m_R}{m_T}.
\fin
Then we get $U_{\nu}$ from the
diagonalization of $M_{\nu}^{I}+M_{\nu}^{II}$, and again $U_e$ from the
diagonalization of $M_e$.

The effect of $M_{\nu}^{II}$ is that to decrease the mixings \cite{rod2}
in such a way that, with the contribution from $U_e$, for a certain range
of $k$, $U_{12}$ falls again within the experimental range
(the impact on $U_{23}$ and $U_{13}$ of our triplet term is almost negligible).
The range of $k$ we found is in fact
\ini
0.08<k<0.18.
\fin
For larger $k$, $U_{12}$ is too small, and for smaller $k$, $U_{12}$ is too large.
We predict $\vartheta_{23}$ nearly maximal, and
$\vartheta_{13} \simeq  \theta_{12}/3 \sqrt2 \simeq 0.05$,
which can be checked by future experiments. 

\section{Conclusion}

We have proposed that a contribution from the triplet term in the type II seesaw
mechanism is important to reconcile the observed quark-lepton complementarity (3)
with the realistic quark-lepton symmetry. We have assumed that the type I term
gives the bimaximal neutrino mixing and the triplet term is proportional to
the right-handed neutrino mass matrix. This framework is well compatible with
the unified $SO(10)$ model. Our study is in some sense the opposite of the one
performed in Ref.\cite{rod1}, where the bimaximal mixing comes from the type II term.
Also other choices of the two seesaw terms have been considered, see for example
Ref.\cite{guo}. Finally we note that the esplicit inclusion of phases in the
charged lepton mass matrix has an impact on the quark-lepton complementarity \cite{ank},
thus the contribution of the triplet term may be reduced or enhanced by this presence.

\end{document}